\documentclass[aps,pra,twocolumn,superscriptaddress]{revtex4-2}
\usepackage{amsmath}    
\usepackage{graphicx}
\usepackage{booktabs}
\usepackage{xcolor}
\usepackage{amssymb}
\definecolor{darkblue}{HTML}{3771C8}
\usepackage[%
    colorlinks=true,
    pdfborder={0 0 0},
    linkcolor=darkblue,
    citecolor = darkblue, 
    urlcolor = darkblue
]{hyperref}
\usepackage[capitalize,nameinlink,noabbrev]{cleveref}

\newcommand{\an}{\hat{a}}
\newcommand{\gqp}{\Gamma_{\text{qp}}}
\newcommand{\nqp}{n_{\text{qp}}}

\begin{document}


\title{Strong kinetic-inductance Kerr nonlinearity with titanium nitride nanowires}

\author{Chaitali Joshi}
\affiliation{Moore Laboratories of Electrical Engineering, California Institute of Technology, Pasadena, CA }
\affiliation{Institute of Quantum Information and Matter, California Institute of Technology, Pasadena, CA}
\author{Wenyuan Chen}
\affiliation{Moore Laboratories of Electrical Engineering, California Institute of Technology, Pasadena, CA }
\author{Henry G. LeDuc}
\affiliation{Jet Propulsion Laboratory, Pasadena, CA}
\author{Peter K. Day}
\affiliation{Jet Propulsion Laboratory, Pasadena, CA}
\author{Mohammad Mirhosseini}
\email{mohmir@caltech.edu}
\homepage{http://qubit.caltech.edu}
\affiliation{Moore Laboratories of Electrical Engineering, California Institute of Technology, Pasadena, CA }
\affiliation{Institute of Quantum Information and Matter, California Institute of Technology, Pasadena, CA}


\date{\today}

\begin{abstract}
Thin films of disordered superconductors such as titanium nitride (TiN) exhibit large kinetic inductance (KI), high critical temperature, and large quality factors at the single-photon level. KI nonlinearity can be exploited as an alternative to Josephson junctions for creating novel nonlinear quantum devices with the potential to operate at higher frequencies and at elevated temperatures. We study a means of magnifying KI nonlinearity by confining the current density of resonant electromagnetic modes in nanowires with a small volume $V \simeq 10^{-4}\text{um}^3$. Using this concept, we realize microwave-frequency Kerr cavities with a maximum Kerr-shift per photon of $K/2\pi = 123.5 \pm 3$ kHz and report a nonlinearity-to-linewidth ratio $K/\gamma = 21\%$. With improved design, our devices are expected to approach the regime of strong quantum nonlinearity in the millimeter-wave spectrum. 


\end{abstract}


\maketitle

\section{Introduction}

Circuit quantum electrodynamics (cQED) systems rely on superconducting qubits based on Josephson junctions. The inductance of a Josephson junction is a nonlinear function of the current tunneling through it, which formally behaves as a Kerr nonlinearity and gives rise to discrete energy levels in a qubit. Alternatively, a current-dependent inductance can be realized via the kinetic energy of current-carrying cooper pairs in a superconducting wire. This source of nonlinear electromagnetic response is currently explored for parametric quantum frequency conversion and amplification in the millimeter-wave band (100-300 GHz), and at elevated ($>1$ K) temperatures \cite{anferov2020, pechal2017}, where using (Al-based) Josephson junctions is not feasible.

While parametric operations are possible with weak nonlinearities, an interesting possibility is to create a strongly nonlinear response where the induced self-Kerr shift per photon in a resonator is comparable to its intrinsic linewidth. Resonators in this regime can be used for synthesizing quantum states of radiation \cite{10.1126/sciadv.abj1916}. Ultimately, when the nonlinearity exceeds the dissipation rate in a resonator the system effectively behaves as qubit \cite{10.1103/physrevlett.79.1467,10.1109/tasc.2021.3065304}. This concept was recently materialized in transmon qubits based on  KI in oxidized (granular) aluminum films (GrAl) \cite{winkel2020a,10.1038/s41535-020-0220-x}. However, the relatively low critical temperature of GrAl (2-3 K) makes it unsuitable for application in the millimeter-wave (mm-wave) band or at elevated temperatures.

Thin films of disordered superconductors such as titanium nitride (TiN), niobium titanium nitride (NbTiN), and niobium nitride (NbN) provide an alternative choice of materials. These films exhibit large KI, and high critical temperature (up to 14.5 K for NbTiN) \cite{zmuidzinas2012}, making them suitable candidates for developing quantum circuits operating at increased temperatures and frequencies while circumventing dissipation due to quasiparticle generation. Additionally, resonators fabricated from these materials can achieve large intrinsic quality factors factors ($10^5-10^7$ in the single-photon regime \cite{leduc2010, vissers2010, coumou2013, sage2011}), and have been widely utilized for high-impedance quantum circuits, amplifiers, and detectors \cite{niepce2019, grunhaupt2019, kamenov2020,landig2018a, samkharadze2016a,hoeom2012, swenson2013a, zmuidzinas2004, malnou2021}. Despite these attractive properties, the KI nonlinearity in nitride films tends to be weaker than GrAl. This is, in part, due to the distinct underlying sources of KI in homogeneously disordered \cite{coumou2015, 10.1103/physrevlett.101.157006}, and granular materials \cite{pracht2016, maleeva2018}, where in the latter the insulating oxide barriers in the microstructure of the film act as weak Josephson links. 


\begin{figure*}[t]
    \centering
\includegraphics[width=0.9\textwidth]{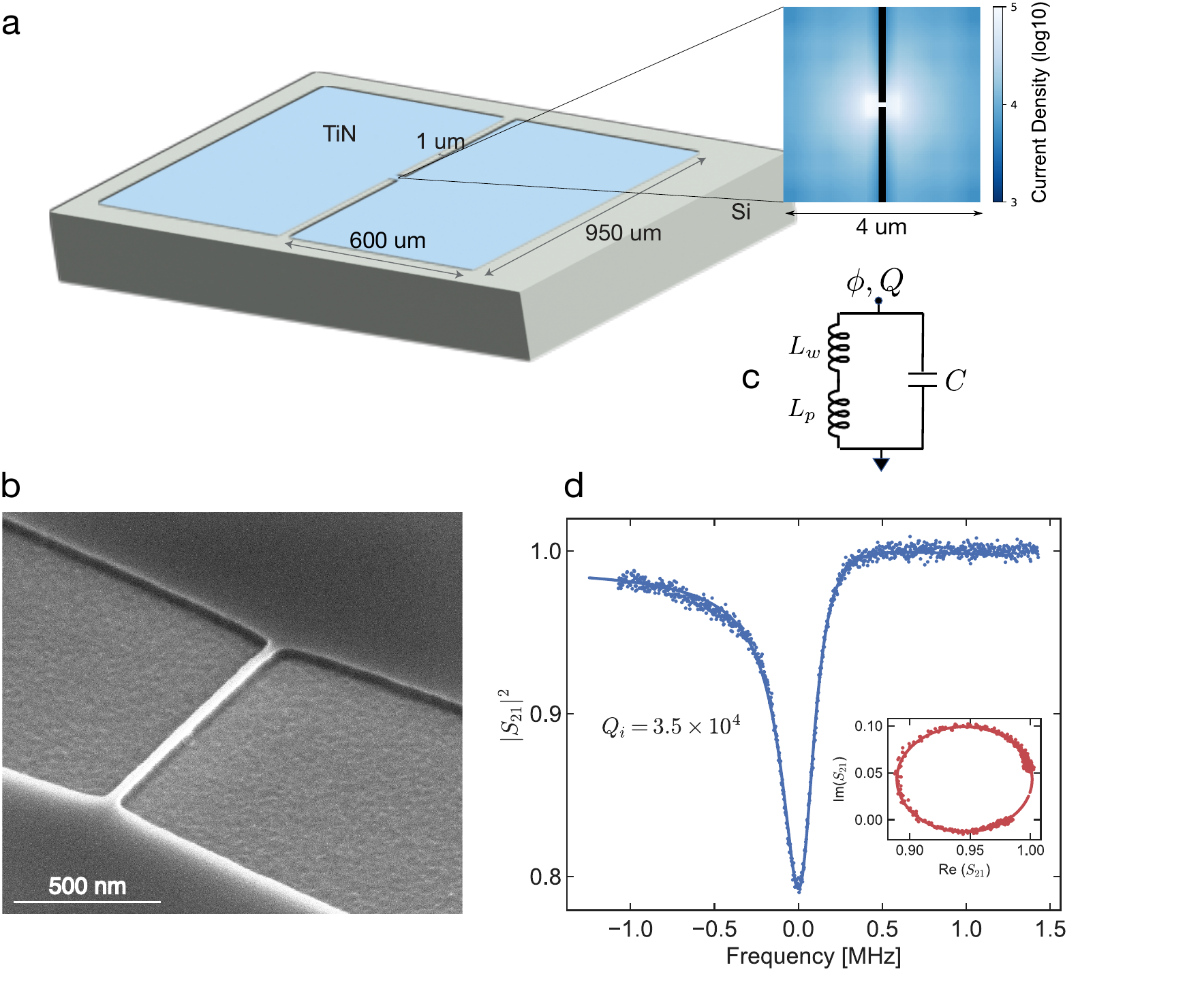}  
\caption{\textbf{Nanowire resonator design:} a) The resonator consists of a miniature nanowire with a widths 20 to 40 nm, shunted by coplanar capacitor pads fabricated from thin film TiN. Inset: Simulated current density in the nanowire region. The current density is maximum in the nanowire and drops in the capacitor pad region in the immediate vicinity of the wire. The nanowire geometry enhances zero-point fluctuations, maximizing the Kerr-shift per photon.  b) Scanning electron microscope (SEM) image of a fabricated nanowire with width $w$ = 38 nm. c) Circuit model of the nanowire resonator. d) Typical linear response of a nanowire resonator (see \cref{tab:results}).}
    \label{fig:design}
\end{figure*}

In this work, we amplify the KI nonlinearity by engineering the geometry of planar resonators fabricated from titanium nitride (TiN). Our design maximizes the nonlinear response by confining the current density of the resonator mode in nanowires with a small lateral width ($\sim$20-200 nm) patterned on a thin film (14 nm thickness). Further, we maximize the magnitude of current quantum zero-point fluctuations by minimizing the resonator impedance, which is achieved by employing narrow-gap planar capacitors. Using these strategies, we achieve self-Kerr nonlinearity rates in the range of $2-120$ KHz in nanowire resonators with intrinsic quality factors reaching $3.5\times {10}^4$ in the best devices. The maximum measured single-photon Kerr-shift to linewidth ratio in our experiment reaches 0.21, which significantly exceeds the nonlinearities reported in prior work with homogeneously disordered superconductors \cite{hoeom2012}. These results motivate future work in optimizing device design, where quantum nonlinearities are expected to be within reach at higher frequencies.

 \begin{figure*}[t]
\centering
\includegraphics[width=1\textwidth]{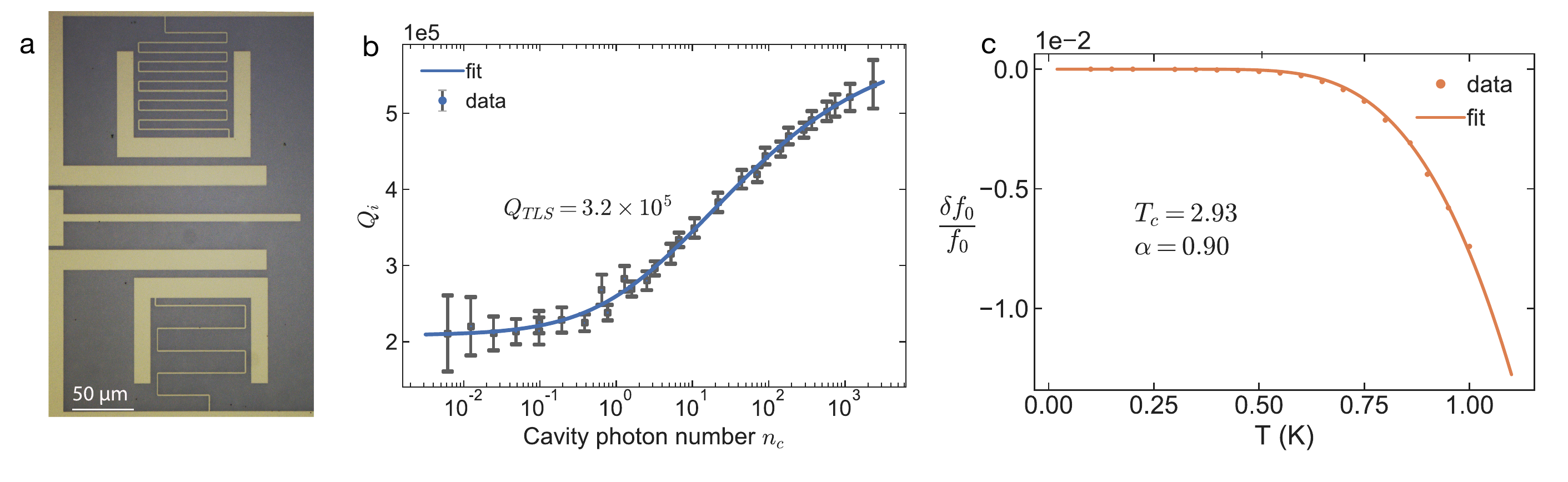}    
\caption{\textbf{Film characterization:} a) Optical image of two compact, quarter-wave test resonators. The resonator consists of a meandered wire shunted by a C-shaped capacitor pad. b) Measured intrinsic quality factor of the compact resonators as a function of mean cavity photon number $n_c$, showing characteristic behavior of TLS loss. The power dependent quality factor $Q_\mathrm{TLS}$ is $3.2 \times 10^5$. At powers exceeding $n_c = 3\times 10^3$, the resonator response becomes nonlinear. c) Measurement of critical temperature $T_c$ using temperature-dependent resonant-frequency shift. Fits to data are from numerical integration of surface-impedance obtained from the Mattis-Bardeen equations.}
    \label{fig:film_ch}
\end{figure*}

\section{Device Design}
Our design concept is to realize an electromagnetic resonant mode with the magnetic energy density confined to a small volume with a large current density. This can be achieved in a compact microwave resonator, where a nanowire with a large KI connects the two pads of a planar rectangular capacitor (see \cref{fig:design}a). For weak currents, the KI of the nanowire can be modeled as a quadratic function $L_w/L_0 =  1 + \left(I/I_\ast \right)^2$, where $I_\ast$ is the nonlinear scaling current set by material properties and wire cross-section \cite{zmuidzinas2012}. The Hamiltonian of this $LC$-resonator can be written as 
\begin{align}
    {H} = \frac{Q^2}{2C} + \frac{\phi^2}{2L} + \frac{1}{4}\frac{\phi^4}{I_\ast^2} \frac{L_w}{L^4 }. 
    \label{eq:h1}
\end{align}
Here, $Q$ and $\phi$ are the charge and flux node variables, and $C$ is the total capacitance. The parameter $L$ is the total inductance of the resonator, which includes the KI of the nanowire ($L_w$) as well as the contribution from the non-zero current distribution in the capacitor pads ($L_p$). The first two terms in the Hamiltonian describe a harmonic oscillator with frequency $\omega = 1/\sqrt{LC}$ and the last term models the nonlinear behavior due to the current-dependent KI in the nanowire. Applying circuit quantization, \cite{10.1109/tasc.2021.3065304, 10.1103/prxquantum.2.040204}, \cref{eq:h1} can be rewritten in terms of the creation ($\an^\dagger$) and annihilation ($\an$) operators of the LC circuit. The nonlinear term in the Hamiltonian can be simplified by expanding the quadratic current term as $\hat{I}^4 = I_\mathrm{ZPF}^4 (\an + \an^\dagger)^4$, where $I_\mathrm{ZPF} = \sqrt{\hbar\omega/2L}$ is the current zero-point fluctuation. Retaining only the photon-number conserving terms, we find 
\begin{align}
    \hat{H} & = \hbar (\omega - K)\left(\an^\dagger \an + \frac{1}{2}\right) - \frac{K}{2}\an^\dagger \an^\dagger \an \an  \nonumber, 
\end{align}
with the Kerr-shift per photon $K$ defined as
\begin{align}
     K  = \frac{3}{2}\omega \frac{L_w}{L} \left(\frac{I_\mathrm{ZPF}}{I_\ast}\right)^2. 
     \label{eq:kshift}
\end{align}

 
As evident in \cref{eq:kshift}, a unity Kerr-shift to linewidth ratio ($K/\gamma \approx 1$) can be achieved if the zero-point fluctuations of the current are enhanced sufficiently to reach $(I_\mathrm{ZPF}/I_\ast)^2 \approx 1/Q$, where $Q = \omega/\gamma$ is the quality factor of the resonator. We pursue this goal by minimizing the total device inductance to achieve a large $I_\mathrm{ZPF}$ while maintaining a sufficiently large wire participation ratio $L_w/L$. 

The expression for the Kerr coefficient can be further simplified by substituting the scaling current as a function of material properties \cite{zmuidzinas2012}
\begin{align}
    I_\ast = J_\ast wt = \sqrt{\frac{ \pi N_0 \Delta_0^3}{\hbar \rho_n} } wt,
\end{align}
where $N_0$ is the density of electron states at the Fermi energy and  $\Delta_0 \approx 1.76 k_BT_c$ is the energy gap at zero temperature. The parameter $\rho_n$ denotes the normal-state resistivity of the film,  $w$ is the wire width, and $t$ is the film thickness. Similarly, the kinetic inductance of the nanowire can be expressed as a  function of geometry and material properties $L_w = \hbar (\rho_n/\pi\Delta_0) (l/ wt)$, where $l$ is the wire length. Finally, defining the participation ratio of KI in the wire to the total inductance as $\alpha = L_w/L$, \cref{eq:kshift} can be rewritten as 
\begin{align}
    K = \frac{3}{4}\frac{\hbar \omega^2}{N_0 \Delta_0^2} \frac{\alpha^2}{V},
    \label{eq:kshift_V}
\end{align}
where $V = w \cdot l \cdot t$ is the volume of the nanowire.

In terms of the material properties, an important observation is that the Kerr coefficient is independent of the film resistivity (so far the nanowire's KI dominates all sources of parasitic inductance). Additionally, the inverse scaling with the energy gap $\Delta_0$ motivates our choice of TiN over niobium-based nitride films, which in general have higher critical temperatures. The geometrical scaling is by and large set by the parameter $V$, which favors nanowires with the minimum possible dimensions. However, in the presence of a non-zero pad inductance $L_p$, reducing the wire length beyond a certain point reduces the Kerr coefficient via the participation ratio $\alpha$. Considering this trade-off, choosing optimum dimension parameters requires the knowledge of film resistivity and the feasible fabrication geometries for nanowires. In our experiment, we realize $L_w$ in the range of 800 pH - 1.6 nH, requiring a shunt capacitance on the order of 0.3 pF to achieve resonance frequencies in our measurement band (6-8.5 GHz). 

Realizing a capacitance of this magnitude with planar geometries is challenging. A large capacitor footprint translates to increased radiation loss and the formation of parasitic self-resonances. Alternatively, choosing narrow gaps and long aspect ratios in interdigitated geometries results in additional loss from surface two-level defects and a reduced participation ratio due to increased pad inductance $L_p$. Using finite-element method (FEM) modeling as a guide to bound these effects, we have chosen a symmetric rectangular capacitor geometry with a total pad area of $1.14~ \mathrm{mm}^2$ and a narrow gap of $1 ~\mu\mathrm{m}$ (see \cref{fig:design}a). As detailed in the following sections, we perform experiments on fabricated devices to evaluate the range of quality factors within reach for this extreme geometry.

  \begin{figure*}[t]
    \centering
    \includegraphics[width = 1.05\textwidth]{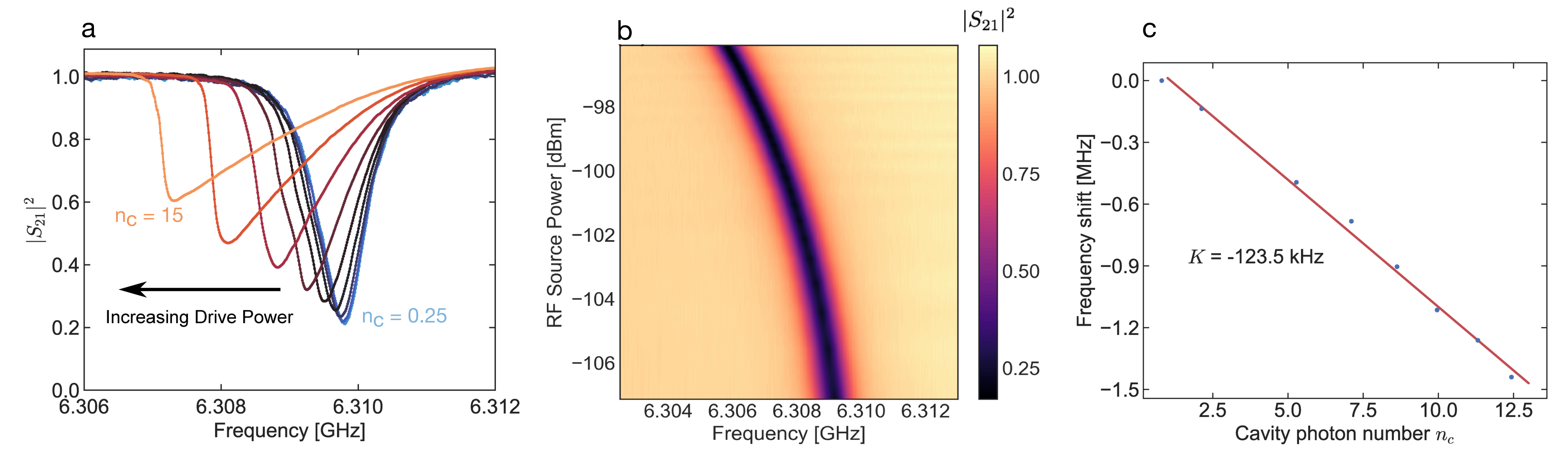}
    \caption{\textbf{Nonlinear response of a nanowire resonator:} Measured nonlinear response of a nanowire resonator with resonance frequency $f_0$ = 6.3 GHz, wire width $w$ = 18 $\pm$ 2 nm and wire length $l$ = 460 nm. a) Self-Kerr response measured at increasing drive powers. The resonator shows characteristic Duffing-oscillator-type behavior. The intracavity photon number is swept from 0.25 to 15, with equal steps on a log scale. (b) Two-tone spectroscopy in the few-photon regime. A radio-frequency drive (blue-detuned by 5.3 MHz from the resonance) is used to drive the resonator and a weak probe is used to measure the coherent response. (c) Extracted Kerr-shift per photon ($K/2\pi$ = 123.5 kHz) from the two-tone spectroscopy measurement.}
    \label{fig:Kerr}
\end{figure*}

 \section{Film characterization}
 \label{sec:film}
Our goal of maximizing nonlinearity motivates using films with minimum possible thicknesses. However, reducing the film thickness results in increasing disorder, which eventually induces a superconducting-to-insulating transition \cite{leduc2010,shearrow2018}. As a result, the choice of film thickness requires balancing the competing requirements for achieving a large kinetic inductance, a large critical temperature, and a low intrinsic loss tangent. Considering these factors and previous results in the literature \cite{leduc2010} we choose to work with films in the 10-20 nm thickness range. The TiN samples used in our experiment were deposited using reactive sputtering on a high-resistivity ($\rho > 10 ~\mathrm{k} \Omega$) Si substrate and measured with an ellipsometer to have a thickness of 14 nm.

To extract the film parameters, we fabricate and measure quarter-wave `test' resonators using the standard single electron-beam lithography followed by ICP/RIE etching in SF$_6$/Ar (\cref{fig:film_ch}a).
We cool samples in a dilution refrigerator with a base temperature of 10 mK and measure coherent reflection from the devices via a circulator using a vector network analyzer (VNA). 
We extract the resonance frequency and the internal quality factors of the resonators by fitting a Lorentzian lineshape to the measured resonator response (see \cref{fig:design}c). We fit the extracted resonance frequency for resonators with varying wire lengths to FEM simulation results to calculate the film sheet KI ($L_s = 40\pm 1$ pH/$\square$, see \cref{app:sheet}). Due to the large wire-cross sections used in these experiments (0.25-2 $\mu$m), the test resonators exhibit weak nonlinearities ($K/2\pi\approx$ 5-50 Hz), which is nevertheless observable at high photon numbers. Characterizing the nonlinearity using using two-tone spectroscopy (detailed in the next section), we find the nonlinear scaling current density of the film ($J_\ast = 3.95 \pm 0.75 \text{ MA}/\text{cm}^2$).  

We characterize the film loss tangent by fitting the change in the internal quality factor of the resonators as a function of the number of photons in the cavity. As evident in \cref{fig:film_ch}b, the internal $Q$-factors shows a saturation behavior that is characteristic of TLS loss \cite{vissers2010, sage2011}. We perform $Q$-factors measurements over a sample size of seven test resonators and find an average (standard deviation) of $2.1\times{10}^5$ ($9\times{10}^4$) at sub-single-photon powers. The $Q$-factors at high photon numbers do not show excessive quasi-particle loss despite the thin film thickness, and are comparable to reported values in previous work with films deposited via atomic layer deposition (ALD) \cite{shearrow2018}. Importantly, we did not observe a statistically significant trend with respect to the wire width of the test resonators.

Finally, we measure the critical temperature of the TiN film by measuring the temperature dependence of the compact resonator frequency (\cref{fig:film_ch}c). The temperature-dependence of the surface impedance $X_s$ results in a change in the resonator frequency, which is given by \cite{gao, klemencic}, 
\begin{align}
    \frac{\delta f_0}{f_0} = -\frac{\alpha}{2} \frac{X_s(T) - X_s(0)}{X_s(0)}
\end{align}
where $\alpha$ is the KI inductance fraction of the resonator (extracted from FEM simulation of the device) and $\delta f_0$ is the change in the cavity resonance frequency. We use surface impedance calculated from numerical integration of the Mattis-Bardeen equations (\cite{gao, garrett2021}) to fit the fractional change in resonance frequency $\delta f_0/f_0$ and extract $T_c = 2.9$ K.

\section{Nonlinear response of nanowire resonators}

Having established our film parameters, we aim to realize nanowire resonators with the smallest feasible dimensions. To achieve this, we have optimized our fabrication procedure and achieved wire widths as small as 20 nm with run-to-run repeatability of about 5 nm. 
We have measured the nonlinear response of fabricated resonators with varying wire dimensions (\cref{tab:results}). The dimensions of the capacitor pads were chosen to be identical for all resonators, and the wire length was adjusted, for each width, to achieve resonance frequencies in the range of 6-8.5 GHz.


We perform linear spectroscopy and find that the nanowire resonators exhibit the characteristic shape of Duffing oscillators (see \cref{fig:Kerr}a), consistent with a Kerr nonlinear response. As expected, beyond a critical circulating power in the resonator, the system undergoes bifurcation and exhibits bistability \cite{eichler2012a}. In addition to the power-dependent frequency shift, we also observe an increase in the internal linewidth of the resonator with the increasing number of photons in the cavity. Given the very small photon numbers used in these measurements ($n_c \sim 1-10$), the broadening is unlikely to be caused by classical nonlinear loss processes. Instead, we attribute this effect to dephasing caused by quantum fluctuations in the number of interactivity photons (i.e., the photon shot noise). This explanation is further substantiated by observing a linear dependence of the linewidth from the fit to the square root of the intracavity photon number (see \cref{app:shotnoise}). Despite the rich underlying physics, which verifies the strong nonlinear response of our system, this photon-number dependent cavity linewidth makes it challenging to extract the self-Kerr coefficient ($K$) from a direct fit to the Duffing oscillator model. 

As an alternative way of finding $K$, we perform two-tone spectroscopy. In these measurements, a detuned coherent pump tone is applied to the cavity while measuring the spectral response to a weak resonant probe (see \cref{fig:Kerr}b). The power of the probe tone is set such that the cavity population from the resonant drive remains two orders of magnitude lower than that of the pump tone. Crucially, the large detuning of the pump tone from the cavity in this measurement scheme leads to a much faster fluctuation time-scale in the number of intracavity photons, effectively eliminating broadening from photon shot noise \cite{gambetta2006}.
As evident in \cref{fig:Kerr}c, varying the number of photons from the pump does not lead to any significant change in the resonator linewidth, further verifying the role of photon-shot noise.

\begin{figure}[t]
    \centering
    \includegraphics[width = 0.48 \textwidth]{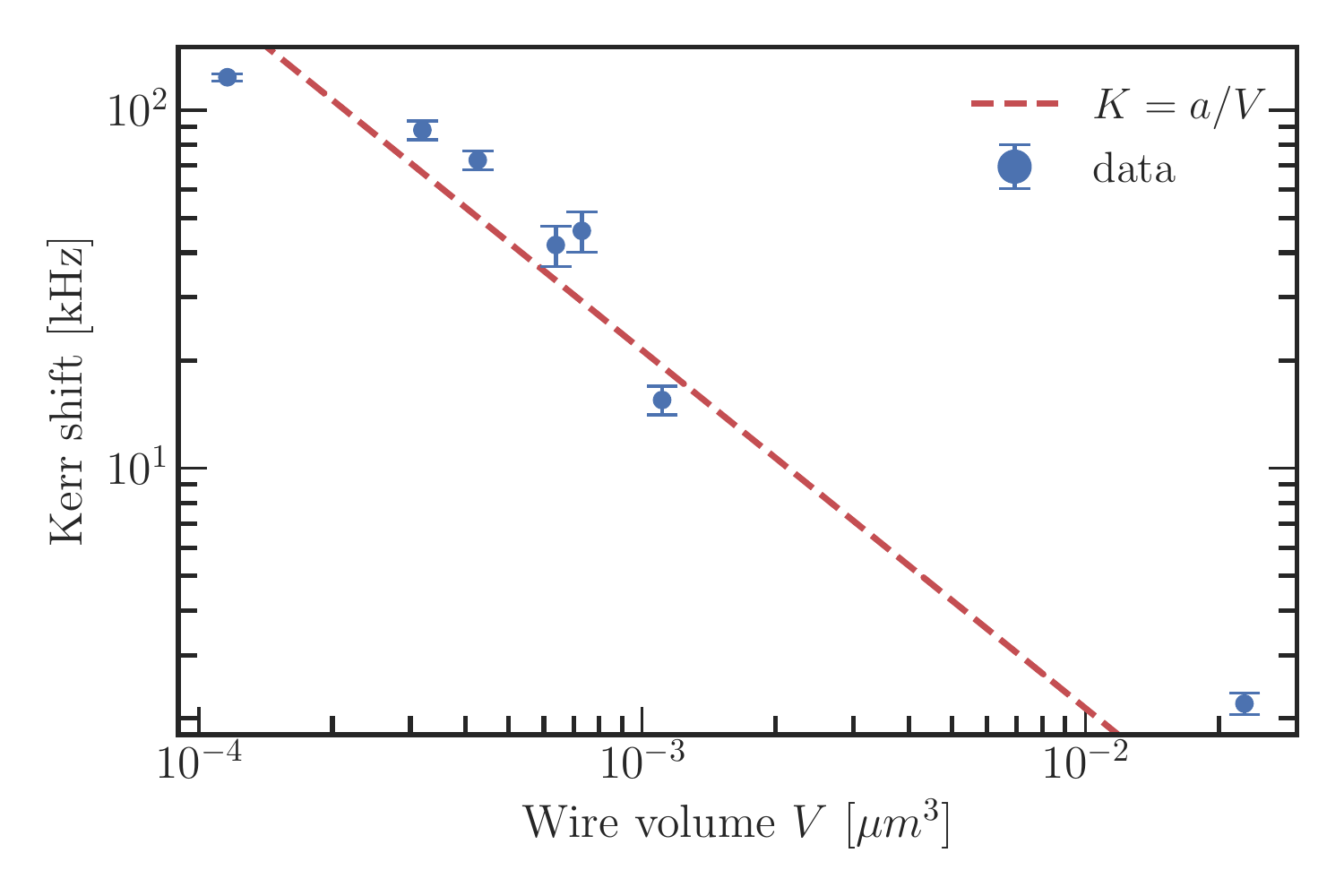}
    \caption{\textbf{Geometric dependence of nanowire nonlinearity:} Measured Kerr-shift for range of wire volumes varying by two orders of magnitude. The largest Kerr-shift is obtained for the wire with smallest mode volume $V$ = 1.2 $\times 10^{-4}$ $\mu \text{m}^3$ and decreases to 2.2 kHz for a volume of $V = 2.3 \times 10^{-2} \mu \text{m}^3$. The red dotted curve is a fit to the model $K = a/V$, with $a = 2 \times 10^{-2}$ \text{kHz}-$\mu \mathrm{m}^3$. }
    \label{fig:nonlinearity_vs_width}
\end{figure}

To fit the data, we calibrate the power delivered to the chip precisely by measuring the frequency-dependent input line attenuation using thermometry of a cryogenic 50 $\Omega$ termination (see \cref{app:line_attenuation}). From linear fits to the cavity-frequency shift, we extract the self-Kerr coefficient of each resonator (see \cref{fig:Kerr}c). The measurement results are summarized in \cref{tab:results}. We find Kerr-shift to linewidth ($K/\gamma$) ratios in a range of 0.7\% to 21\%, and internal $Q$-factors in the range of $0.2 - 3.5 \times 10^4$ (\cref{tab:results}). The measured self-Kerr shift is largest for the wire with smallest mode volume $V = 1.2 \times 10^{-4} \mu \mathrm{m}^3$. After that, we observe a monotonic decrease in Kerr-shift for increasing wire volume, in agreement with the expected $1/V$ scaling behavior (see \cref{fig:nonlinearity_vs_width}). Using a fit to the model (\cref{eq:kshift_V}), we extract participation ratios $\alpha$ in the range of 0.7 - 0.97. These participation ratios are larger than those expected using our measured value of sheet inductance ($L_s$) and the wire dimensions obtained from imaging the devices. We attribute this discrepancy to contributions to the nonlinear response from the current distribution in the capacitor pads, which are not included in our lumped-element model.

We find the internal $Q$-factors of the nanowires resonators to be smaller (by about an order of magnitude) than those found in lumped-element test resonators used for film characterization. Importantly, we do not observe significant wire-width dependence for the internal $Q$-factor; the device with the largest nonlinearity and narrowest wire width has an internal $Q$-factors of 1.08 $\times 10^4$. We have verified that the measured $Q$-factors are dominated by energy loss, despite the presence of a slow dephasing mechanism (see \cref{app:dephasing}). Below, we discuss the dominant loss mechanisms in our devices and potential avenues for improving the $K/\gamma$ ratio. 

{
\renewcommand{\arraystretch}{1}
\begin{table}
\centering
\resizebox{1\linewidth}{!}{

\begin{tabular}{lccccccc}
\toprule
 & $f_0$ & Width & Length & Linewidth & Intrinsic & Kerr shift & $K/\gamma$ \\ 
& (GHz) & (nm) & (nm) & ($\gamma$, kHz) & $Q$-factors & ($K/2\pi$, kHz) &  ($\%$)\\
\midrule
1 & 6.30  & 18 $\pm$ 2 & 460 & 580 & $1.08 \times 10^4$ & 123.5 $\pm$ 3 & 21 \\
2 & 8.48$^{\textcolor{blue}{\ast}}$ &	38 $\pm$ 2 & 600 & 	1600 & $0.53 \times 10^4$ & 88 $\pm$ 5.3 & 5.5\\
3 & 8.07 & 38 $\pm$ 2 & 800 & 800 & $1 \times 10^4$ & 72.5 $\pm$ 4.5 & 9 \\
4 & 7.03 & 36 $\pm$2 & 1450 & 310 & $2.3 \times 10^4$ & 46 $\pm$ 6 & 15  \\
5 & 7.12$^{\textcolor{blue}{\ast}}$ & 38 $\pm$ 2 & 1200 & 	1000 &  0.71 $\times 10^4$ & 42 $\pm$ 5.4 & 4.2 \\ 
6& 6.50 & 44 $\pm$	2 & 1800 & 2300 & $0.2 \times 10^4$ & 15.5 $\pm$ 1.5 & 0.7 \\   
7& 7.70 & 225 $\pm$ 2 & 7250 & 220 & $3.5 \times 10^4$ & 2.2 $\pm$ 0.15 & 1\\
\bottomrule
\end{tabular}
}             
\caption{\textbf{Summary of results.} The uncertainty in calculating the Kerr coefficient is dominated by the uncertainties in the measurement of the frequency-dependent input line attenuation. The internal linewidth $\gamma$ of all resonators was extracted from the complex amplitude response at single-photon ($n_c \approx 0.3$) drive levels. $^{\textcolor{blue}{\ast}}$ These devices were on a separate chip (see text).}
\label{tab:results}
\end{table}
}
%

\section{Dissipation mechanisms}

The narrow-gap capacitor geometry and the resulting large electric fields in our designs make our devices sensitive to TLS loss at interfaces \cite{melville2020, amin2022}. Common techniques for extracting the TLS-induced loss involves saturating the TLS bath via by a coherent external drive or by elevating the measurement temperature. However, in our current experiment the small bistability threshold set by the strongly nonlinear response of the cavities does not allow for directly applying these techniques. Additionally, while these measurements are, in principle, possible in test structures with identical electric field distributions but with a diluted nonlinearity (e.g. via wider wires), the inevitable reduction in kinetic inductance leads to much higher resonance frequencies beyond our setup's measurement band. Beyond TLS loss, we expect two primary sources of dissipation in our devices. The large KI fraction of the nanowire resonators makes them susceptible to dissipation from excess quasiparticles generated by stray radiation \cite{grunhaupt2019, mcrae2020, kreikebaum2016, siddiqi2021}. In addition, the large size of the capacitor pads together with a small gap results in a large effective dipole, making the resonators susceptible to radiative loss. Here, we provide order-of-magnitude estimates for loss contributions from these sources. 


The rate of dissipation due to a nonequilibium quasiparticle density $\nqp$ can be predicted from the Mattis-Bardeen equations for the complex conductivity, and is given by \cite{barends2011a}, 
\begin{align}
    \gqp = \frac{\alpha \omega_0}{\pi} \sqrt{\frac{2\Delta_0}{\hbar \omega_0}} \frac{\nqp}{D(E_F)\Delta_0}
    \label{eq:qp_loss}
\end{align}
where $\alpha$ is the kinetic inductance fraction, $\omega_0$ is the resonance frequency, $\nqp$ is the excess quasiparticle density, and $D(E_F)$ is the density of Cooper pairs at Fermi energy. We evaluate \cref{eq:qp_loss} using parameters reported in literature for TiN thin films ($D(E_F) = 2 \times 10^{10}/\mu \text{m}^3-eV$ \cite{gao2012}, $\nqp = 10/\mu \text{m}^3$ \cite{barends2011a, siddiqi2021}), our measured critical temperature, and the simulated KI fraction $\alpha $.  As a verification of the calculation, we find the QP-limited $Q$-factors of the test resonators used for film characterization (in \cref{sec:film}) ($4.8 \times 10^5$), which is found to be in close agreement with measured $Q$-factors at high powers ($2.1 \times 10^5$). For the nanowire resonators, the calculation gives a similar decay rate in the range of 12-14 kHz, which is much smaller than the observed values in the experiment. Interestingly, the calculated quasi-particle limited decay rates are about an order of magnitude smaller than the largest measured single-photon Kerr shifts, pointing to the intrinsically low-loss nature of KI as a source of nonlinearity.

We estimate radiation loss contributions by modeling the near-field coupling of the nanowire resonators to the packaging box modes. Our device is packaged within a copper box with air-gaps below and above the chip, with the lowest order resonant mode located near $f_b$ = 17 GHz. We use a conservative $Q$-factors for the microwave package mode ($\kappa/2\pi = 35$ MHz) \cite{huang2020a} and use analytical expressions for the electric field of coplanar capacitor plates to estimate an upper bound for radiative loss rate of $\Gamma/2\pi \approx 60$ kHz (see \cref{app:Radloss}). While significant, this source of loss does not adequately account for the observed linewidths. Ruling out QP-dissipation and radiative loss, we suspect the device performance is currently limited by TLS loss. 


\section{Conclusion}
In conclusion, we have investigated the limits of KI Kerr nonlinearity by optimizing the geometry of resonators fabricated from thin-film TiN. 
Confining current density in miniaturized nanowires, we have observed a nonlinearity-to-loss ratio of $K/\gamma =0.21$. While the achieved ratio is not sufficient to make a qubit based on this concept in the microwave frequency band, the quadratic scaling of nonlinearity with frequency (\cref{eq:kshift}) promises future opportunities: At a frequency of 100 GHz, this scaling translates to Kerr-shift coefficients in the range of 20-30 MHz per photon. Intrinsic quality factors as large as $3 \times 10^4$ ($\gamma/2\pi = 3$ MHz) have been previously demonstrated using NbN films at this frequency \cite{anferov2020}. Achieving our scaled Kerr-shift coefficients in devices with comparable quality factors would translate to the observation of a strong quantum nonlinear response. This motivates future studies of loss mechanisms in TiN, and a broader set of disordered superconductors, in the mm-wave band. Additionally, the improved Kerr coefficients in our devices may be exploited for frequency conversion from microwave to mm-wave frequencies, where operation with lower pump powers and stronger nonlinearities can reduce absorption heating. Using the measured and scaled Kerr-shifts at microwave and millimeter-wave frequencies we estimate a cross Kerr shift of $K_\mathrm{uw,mm} = 2\left({K_\mathrm{uw} K_\mathrm{mm}}\right)^{1/2} \approx 2\pi\cdot3$ MHz \cite{minev2020}.  Finally, we note the possibility of further improvements by going beyond planar geometries by using parallel-plate capacitors, where large quality factors ($10^5$) have been previously demonstrated \cite{beyer2017}.

\section*{acknowledgments}
 This work was supported by startup funds from the Caltech EAS division, a Braun trust grant, and the National Science Foundation (grant No. 1733907). C.J. gratefully acknowledges support from the IQIM/AWS Postdoctoral Fellowship. We acknowledge Niv Drucker from Quantum Machines for software support while performing the cavity ringdown measurements.

\bibliography{nanowires}

\clearpage
\appendix
\section{Film characterization: sheet inductance}
\label{app:sheet}
\begin{figure}[h]
    \centering
    \includegraphics[width = 0.5\textwidth]{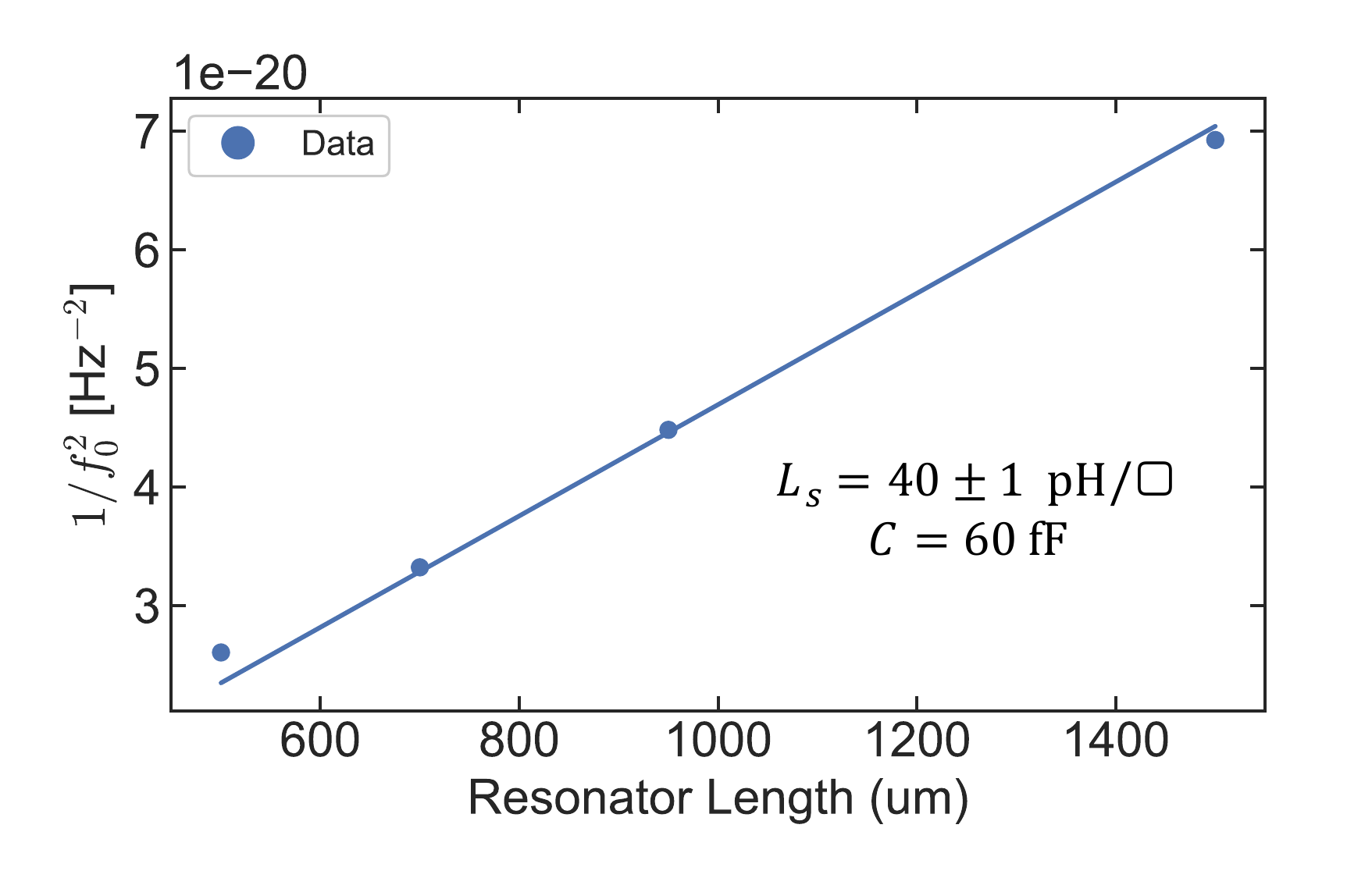}
    \caption{\textbf{Calibration of sheet inductance:} Measured $1/f_0^2$ vs. wire length for quarter-wave resonators with wire width $w= 2 \mu$m. From the slope of the fit $m = 4\pi^2 L_sC/w$ and capacitance extracted from simulations (C = 60 fF), we extract a sheet kinetic inductance of $L_s = 40 \text{pH}/\square$.}
    \label{fig:sheet_inductance}
\end{figure}

We estimate the sheet inductance of the TiN thin films by fitting the extracted resonance frequency of the quarter-wave resonators with wire width $w = 2~ \mu$m and varying wire lengths (\cref{fig:sheet_inductance}). A capacitance of $C = 60$ fF was extracted for the quarter-wave resonator geometry (\cref{fig:film_ch}a) using finite element method simulations. From the slope $m = 4\pi^2 L_sC/w$ of the fit shown in \cref{fig:sheet_inductance}, we obtain a sheet inductance of $L_s = 40 \pm 1$ pH/$\square$. 
{
\renewcommand{\arraystretch}{1}
\begin{table}[h]
\centering
\resizebox{1\linewidth}{!}{

\begin{tabular}{lccc}
\toprule
$f$ & Net Transmission & Gain & Attenuation \\ 
(GHz) & ($S_\text{{IO}}$, dB) & ($G_C$, dB) & ($A_{IN}$, dB)\\
\midrule
4.4 & -5.5 & 68.3 $\pm$ 0.4 & 73.8 $\pm$ 0.4  \\
6.30 & -11.50 & 64.6 $\pm$ 0.1 & 76.1 $\pm$ 0.1 \\
6.56 & -11.95 & 64.5 $\pm$ 0.4 & 76.5 $\pm$ 0.4 \\
7.03 & -13.95 & 63.6 $\pm$ 0.6 & 77.6 $\pm$ 0.6 \\
7.70 & -16.90 & 62.9 $\pm$ 0.3 & 79.8 $\pm$ 0.3 \\
8.07 & -18.30 & 61.4 $\pm$ 0.3 & 79.7 $\pm$ 0.3 \\ 
\bottomrule
\end{tabular}
}
\caption{\textbf{Line attenuation calibration}}
\label{tab:line_attenuaton}
\end{table}
}

\begin{figure}
    \centering
    \includegraphics[width = 0.6\textwidth]{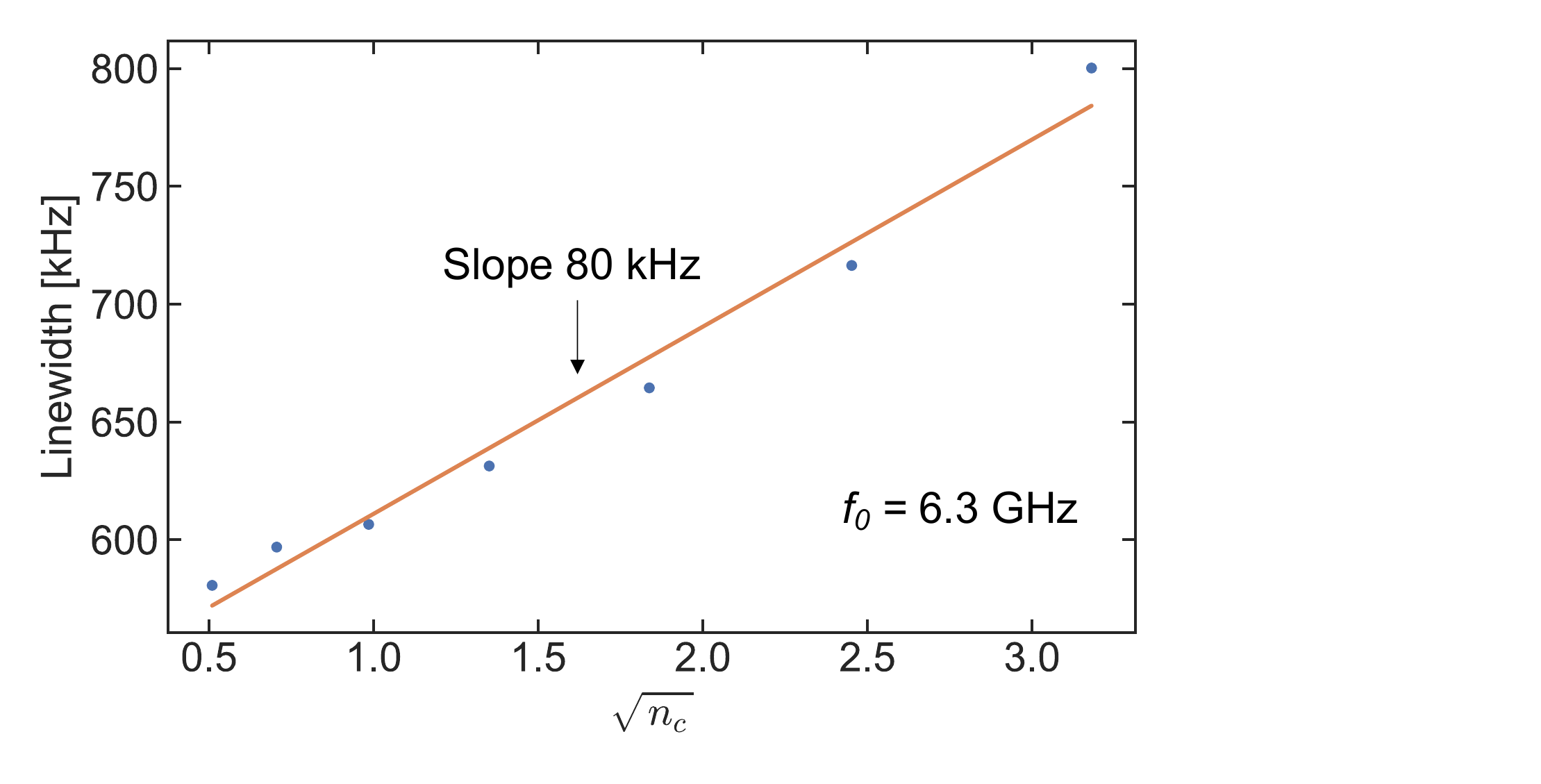}
    \caption{\textbf{Shot-noise induced dephasing in self-Kerr response:} Measured cavity linewdith vs photon number for the self-Kerr data shown in \cref{fig:Kerr}a. The increase in linewidth can be attributed to dephasing due to shot noise in the cavity photon number when the cavity is resonantly driven.}
    \label{fig:shotnoise}
\end{figure}

\section{Line attenuation calibration}
\label{app:line_attenuation}

 \begin{figure*}[t]
    \centering
    \includegraphics[width = 1.05\textwidth]{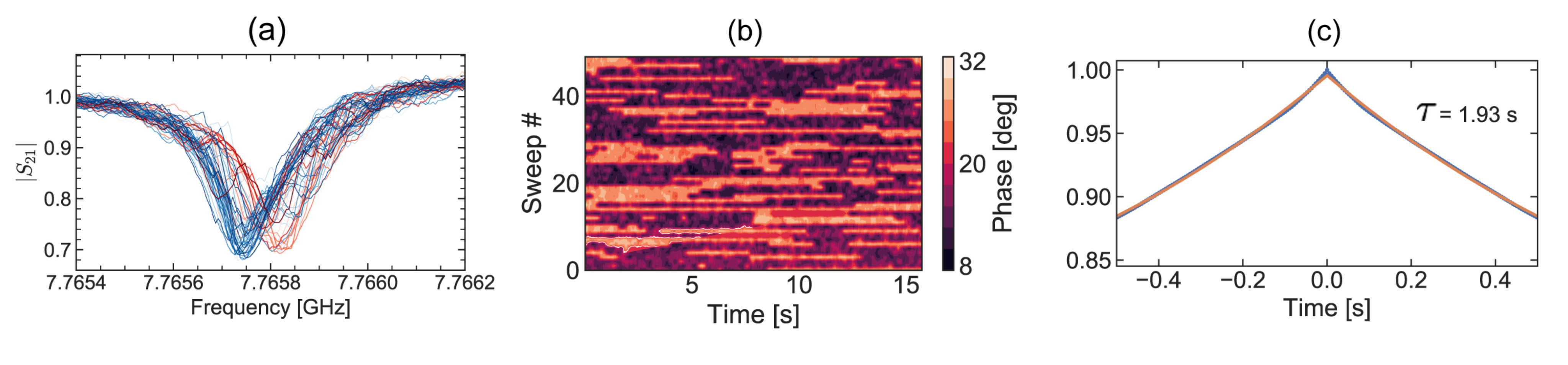}
    \caption{\textbf{Telegraph noise:} a) Single-shot measurements of the frequency response of the resonator with the narrowest linewidth ($f_0$ = 7.7 GHz). The resonator exhibits switching between two states. b) Measured time-dependent phase response, showing characteristics of a telegraph process. c) Auto-correlation of the time-dependent phase of the resonator. We extract a switching time of $\tau$ = 1.93 s.}
    \label{fig:Dephasing}
\end{figure*}

To calculate accurately the power delivered to the chip, we calibrate the frequency-dependent attenuation of our input line using thermometry. The input line contains cryogenic attenuation at each stage of the cryostat to minimize thermal noise conducted from room temperature electronics to the cold stages via the input lines\cite{krinner2019}. In addition, the input line consists of a crogenic circulator (Low noise factory, LNF-CIC4$\_$12A) and a fan-out switch. The output line consists of a HEMT amplifier (LNF-LNC4$\_$8C) thermalized to the 4 K stage and a room temperature amplifier. We first measure the overall frequency-dependent transmission of the input and output line using an RF tone sent from a signal generator. We then calibrate the total gain of the output line using thermometry. A 50 $\Omega$ cryogenic terminator is thermalized to the mixing (MXC) stage of the cryostat. The MXC stage temperature is raised by reducing cooling power by turning off the turbo to reduce $^3$He/$^4$He mixture flow, and applying heat using the MXC stage heater. With no external input power, we measure the output power from the amplifier chain with a spectrum analyzer at different mixing stage temperatures $T_\text{MXC}$. The measured output has contributions from the thermal noise of the resistor at the MXC stage, and the HEMT noise characterized by a fixed noise temperature $T_\text{HEMT}$. The output power measured in an IF bandwidth $\Delta\nu_{IF}$ on the spectrum analyzer is equal to the sum of the Johnson-Nyquist noise from the two sources and is given by, 
\begin{align}
P_{\text{OUT}} = \Delta\nu_{\text{IF}} k_\text{B} G_C \left(T_{\text{MXC}} + T_{\text{HEMT}} \right) 
\label{eq:thermalnoise}
\end{align}

Here, $G_C$ is the net absolute gain factor of the output line from the chip. At $T_\text{MXC} = 10$ mK, the measured output power is dominated by HEMT noise. The output gain $G_C$ can then be calculated by subtracting the contribution of the HEMT noise from the total output power measured at various $T_\text{MXC}$. We perform this measurement at 4 different MXC temperatures between 730 mK and 1.05 K. The mean and standard deviation of the four measurements is used to obtain $G_C$. We use a factor $\eta = (h\nu/kT)/\left(\exp{\left(h\nu/kT\right)} - 1\right)$ to account for corrections to \cref{eq:thermalnoise} due to the Bose-Einstein distribution in the regime $h\nu \sim k_BT$ \cite{nyquist1928}. The input line attenuation can be calculated using, 
\begin{align}
    A_{IN} [\text{dB}] + G_C [\text{dB}] = S_{IO}[\text{dB}]. 
\end{align}
Here, $S_{IO}$ is the net transmission of the input-output lines measured at different frequencies and $G_C [\text{dB}]$ is the calibrated output gain in dB. The results are summarized in \cref{tab:line_attenuaton}. 

\section{Shot-noise induced dephasing}
\label{app:shotnoise}
The measured self-Kerr response of the nanowire resonators shows an increased linewidth when the cavity is populated with few photons (\cref{fig:Kerr}a). The relatively large ratios of Kerr shift to linewidth create the possibility of dephasing due to shot noise when the cavity is populated via a resonant drive. As shown in \cref{fig:shotnoise}, the measured linewidth increases proportional to $\sqrt{n_c}K$. Additionally , we extract a slope of 80 kHz for this linear dependence. which is close to the measured self-Kerr coefficient for this resonator (123.5 kHz). In contrast, we do not observe a significant linewidth broadening in the two-tone spectroscopy measurement, where the cavity is driven off-resonantly (\cref{fig:Kerr}b). In this case, the cavity is only virtually populated, and we estimate the increase in linewidth to be proportional to $nK^2/\Delta$, which is of the order of 10 kHz \cite{gambetta2006}.

\section{Telegraph noise and $T_1$ lifetime measurement}
\label{app:dephasing}
\begin{figure}[t]
    \centering
    \includegraphics[width = 0.45\textwidth]{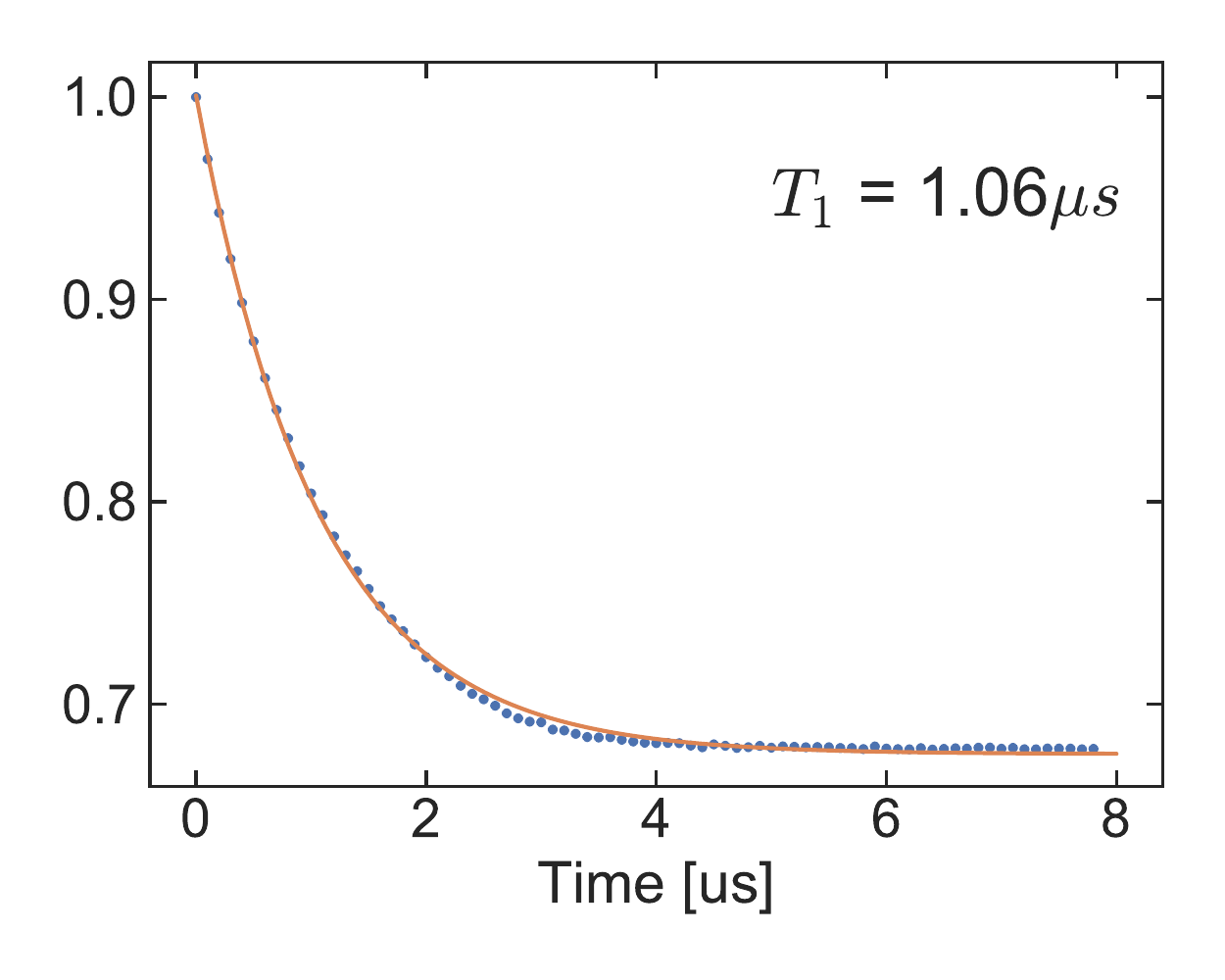}
    \caption{\textbf{Cavity-ring down spectroscopy:} Time-dependent energy decay of the resonator after turning-off of the drive pulse. We measure a $T_1$ lifetime of 1.06 $\mu$s, corresponding to an instantaneous linewidth of 150 kHz for the 7.7 GHz resonator.}
    \label{fig:T1}
\end{figure}
 We observe telegraph noise in the frequency response of the resonator with the narrowest linewidth  (Device \#7, see \cref{tab:results}). We perform single-shot VNA measurements to capture the fast dephasing of this resonator. The resonator exhibits switching between two states, as shown in \cref{fig:Dephasing}a. To measure the characteristic transition time $\tau$ between the two states, we measure the phase response of the resonator continuously using the zero-span mode of the VNA.  As shown in \cref{fig:Dephasing}b, the phase response is reminiscent of the telegraph process. The switching time can be extracted from the autocorrelation function of the telegraph process, given by $\langle \phi(t) \phi(t')\rangle \propto \exp{{-2|t - t'|}/{2\tau}}$, where $\phi(t)$ is the time-dependent phase shown in \cref{fig:Dephasing}b. We extract a switching time of $\tau = 1.93$s. 
 
 \begin{figure*}[t]
    \centering
    \includegraphics[width = 1\textwidth]{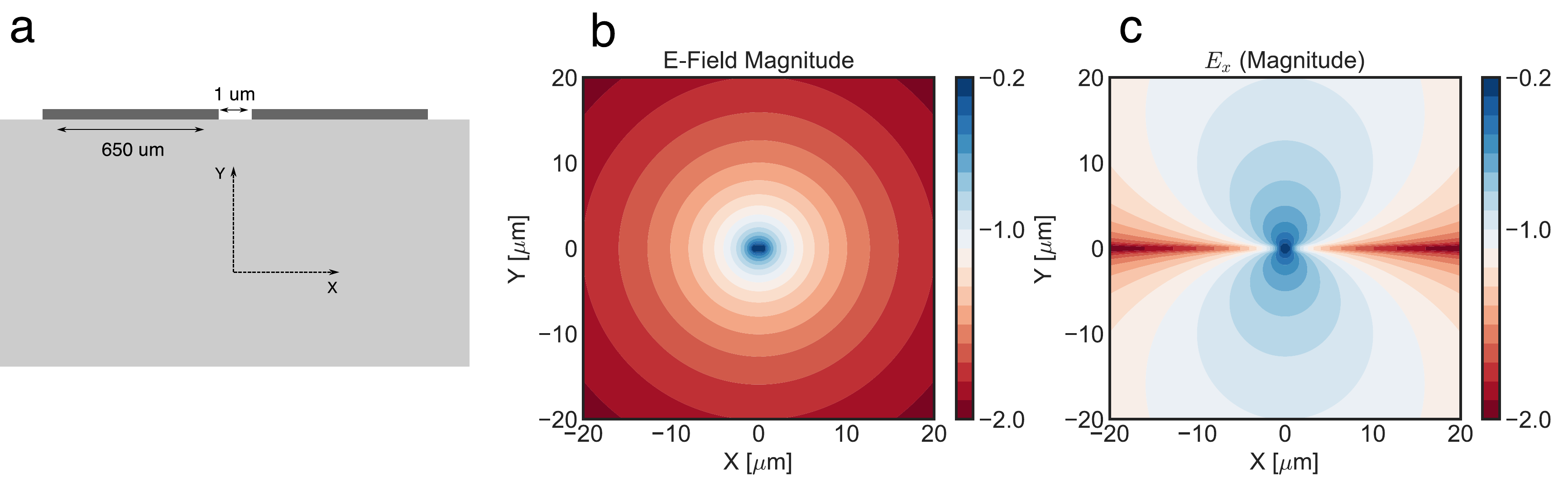}
    \caption{\textbf{E-field distribution of coplanar capacitor pads:} (a) Cross-section of the coplanar capacitor pads. (b) E-field magnitude of coplanar capacitor pads in the vicinity of the capacitor gap, calculated from analytical expressions using conformal mapping. The field distribution resembles that of a dipole. (c) In-plane ($E_x$) field magnitude, which results in the dominant coupling to the TE$_{011}$ box mode. }
    \label{fig:RadiationLoss}
\end{figure*}

 We perform cavity ring-down spectroscopy on this resonator to extract it's $T_1$ lifetime. The resonator was driven using a modulated signal generated from an FPGA, upconverted by frequency mixing with an RF signal generator acting as the local oscillator (LO). The resonator is driven using a $5 \mu s$-long pulse, sufficiently long to ensure that the cavity population is saturated. The drive is then turned off abruptly. The drive power was set to be sufficiently low such that the resonator response is in the linear regime. The time-dependent decay of the cavity population is shown in \cref{fig:T1}. The measured $T_1$ lifetime is 1.06 $\mu$s, corresponding to a decay rate of 150 kHz. Comparing this value with the measured resonator linewidth with long averages (220 kHz), bounds the contributions from dephasing to approximately 70 kHz.

 Telegraph noise has been previously observed in kinetic inductance resonators \cite{zhang2019b, grunhaupt2018, niepce2020, sueur2018}. Potential mechanisms for the origin of this noise are strong coupling to TLS \cite{zhang2019b, niepce2020, sueur2018} or quasiparticle burst noise from high-energy particles \cite{grunhaupt2018}, and requires further investigation which is beyond the scope of this study.

\section{Radiation loss calculation}
\label{app:Radloss}

We estimate the radiation loss of the resonator to the box using using the Purcell decay expression in the far-detuned regime, where the resonator decay rate can be estimated as, 
\begin{align}
    \Gamma =  \frac{\kappa g^2}{\Delta^2}.
    \label{eq:purcell}
\end{align}
 Here, we assume $\Delta \gg \kappa, g$. The parameter $\Delta/2\pi = f_b - f_0$ is the detuning between the resonator and the box mode $f_b$ and $\kappa/2\pi = f_b/Q$ is the decay rate of the box mode. The coupling rate $g$ between the resonator and the box mode is related to the electric field overlap between the cavity and the box modes, and is given by \cite{lin2016, irvine2006}, 
\begin{align}
\hbar g = \sqrt{\dfrac{\hbar \omega_r}{2\epsilon_0 V_r}} \sqrt{\dfrac{\hbar \omega_b}{2\epsilon_0 V_b}} \int dV \epsilon \epsilon_0 |\mathbf{a_r^\ast}\cdot \mathbf{a_b}|
       \label{eq:couplingrate}
\end{align}
where $V_r$ and $V_b$ are the effective mode volumes of the resonator and box mode respectively. The coefficients $\mathbf{a_r}$ and $\mathbf{a_b}$ are normalized, dimensionless amplitudes such that the electric-field of $\mathbf{E_i}$ is given by $\mathbf{E_i} =  \sqrt{{\hbar \omega_i}/{2\epsilon_0 V_i}} \mathbf{ a_i}, i \in [r,b]$. The integral is carried over the volume of the box. The field distribution $\mathbf{E_r}$ of the resonator mode can be approximated from analytical expressions for coplanar capacitor pads obtained using conformal mapping \cite{2015, ramer1982}. The electric field distribution for the resonator mode is shown in \cref{fig:RadiationLoss}. As expected, the field distribution resembles that of a dipole due to the large electric field in the small capacitor gap. Using FEM simulations, the lowest order box mode is found to be a TE$_{011}$ mode at $f_b$ = 17 GHz. Assuming the resonator is placed in the center of the box where the E-field of the box mode peaks, we evaluate \cref{eq:couplingrate}
 numerically and determine a coupling rate $g/2\pi = 410$ MHz. For a typical resonator frequency $f = 7$ GHz and assuming a typical package mode Q $\approx$ 500, we obtain a radiation loss rate of $\Gamma/2\pi = 57$ kHz using \cref{eq:purcell}. The degenerate TE$_{101}$ mode does not couple to the resonator mode due to it's orthogonal polarization. The immediate higher order modes at higher frequencies also do not contribute significantly as they have E-field nodes at the center of the box, and the decay rate is further suppressed due to their large detuning $\Delta$ relative to the resonator mode.

\end{document}